\documentclass[english,nalefd]{achemso}
\usepackage[LGR,T1]{fontenc}
\usepackage[latin9]{inputenc}
\usepackage{textcomp}
\usepackage{amstext}
\usepackage{graphicx}
\usepackage{epstopdf}
\makeatletter

\title{High-efficiency generation of nanoscale single silicon vacancy defect
array in silicon carbide}
\author{Junfeng Wang}
\affiliation{Division of Physics and Applied Physics, School of Physical and Mathematical
Sciences, Nanyang Technological University, Singapore 637371, Singapore}
\author{Yu Zhou}
\affiliation{Division of Physics and Applied Physics, School of Physical and Mathematical
Sciences, Nanyang Technological University, Singapore 637371, Singapore}
\author{Xiaoming Zhang}
\affiliation{Division of Physics and Applied Physics, School of Physical and Mathematical
Sciences, Nanyang Technological University, Singapore 637371, Singapore}
\author{Fucai Liu}
\affiliation{Center for Programmable Materials, School of Materials Science \& Engineering, Nanyang Technological University, 50 Nanyang Avenue, Singapore 639798, Singapore.}
\author{Yan Li}
\affiliation{Key lab of Quantum Information, CAS, University of Science and Technology
of China, Hefei, Anhui, 230026, P. R. China}
\author{Ke Li }
\affiliation{Division of Physics and Applied Physics, School of Physical and Mathematical
Sciences, Nanyang Technological University, Singapore 637371, Singapore}

\author{Zheng Liu}
\affiliation{Center for Programmable Materials, School of Materials Science \& Engineering, Nanyang Technological University, 50 Nanyang Avenue, Singapore 639798, Singapore.}
\author{Guanzhong Wang}\email{gzwang@ustc.edu.cn}
\affiliation{Hefei National Laboratory for Physical Science at Microscale, and
Department of Physics, University of Science and Technology of China,
Hefei, Anhui, 230026, P. R. China}
\author{Weibo Gao}\email{wbgao@ntu.edu.sg}
\affiliation{Division of Physics and Applied Physics, School of Physical and Mathematical
Sciences, Nanyang Technological University, Singapore 637371, Singapore}


\ProvideTextCommand{\~}{LGR}[1]{\char126#1}

\makeatother

\usepackage{babel}
\begin{document}
\begin{abstract}
Color centers in silicon carbide have increasingly attracted attention in recent years owing to their excellent properties such as single photon emission, good photostability, and long spin coherence time even at room temperature. As compared to diamond which is widely used for holding Nitrogen-vacancy centers, SiC has the advantage in terms of large-scale, high-quality and low cost growth, as well as advanced fabrication technique in optoelectronics, leading to the prospects for large scale quantum engineering. In this paper, we report experimental demonstration of the generation of nanoscale $V_{Si}$  single defect array through ion implantation without the need of annealing. $V_{Si}$ defects are generated in pre-determined locations with resolution of tens of nanometers. This can help in integrating $V_{Si}$ defects with the photonic structures which, in turn, can improve the emission and collection efficiency of $V_{Si}$ defects when it is used in spin photonic quantum network. On the other hand, the defects are shallow and they are generated $\sim 40nm$ below the surface which can serve as critical resources in quantum sensing application.
\end{abstract}

Defects in silicon carbide (SiC) stand out in recent years due to their outstanding features, such as high-quality growth, high thermal conductivity, and mature nanofabrication techniques \cite{key-1,key-2,key-3,key-4,key-5,key-6,key-7,key-8,key-9,key-10,key-11,key-12,key-13,key-14,key-15,key-16}. Similar to the nitrogen-vacancy centers in diamond \cite{key-17}, silicon vacancy ($V_{Si}$) and divacancy defects can be used as spin qubits and can be optically polarized and controlled by microwave \cite{key-1,key-2,key-3,key-5,key-9,key-6,key-10,key-13}. For $V_{Si}$ defects, coherent control of single spin has been achieved at room temperature and the coherence time reaches about 160 $\mu s$ \cite{key-3}. Moreover, the coherence time of the $V_{Si}$ defect ensemble has been improved to about 70 ms using dynamics decoupling at cryogenic temperature \cite{key-15}. In particular, $V_{Si}$ defect has been shown to have application in quantum sensing of magnetic field\cite{key-9,key-11} and of temperature \cite{key-9}.  Most of $V_{Si}$ defect related experiments were performed on ensembles \cite{key-9,key-11,key-15,key-16}. In order to study the properties of single $V_{Si}$ defects, efficient generation of single $V_{Si}$ defects is required \cite{key-3,key-4,key-5}. Recently, there are two methods that have been developed to generate single $V_{Si}$ defects: high energy electron irradiation \cite{key-3} and neutron irradiation \cite{key-4}. However, in both methods, the generation efficiency of the single $V_{Si}$ defects is low and the positions of the single $V_{Si}$ defects are randomly distributed.  Towards the application in spin photonics network \cite{key-18}, the creation of the defects in well-defined locations is essential in order to enhance their emission efficiency by integrating them into photonic structures.  This shows that improvement in both the generation efficiency and the position accuracy of single $V_{Si}$ defects is indispensable.

In this work, we experimentally realized the efficient generation of nanoscale single $V_{Si}$ defect array in SiC.  The $V_{Si}$ defects array is created by using 30 keV carbon ion implantation through an array of $65\pm10$ nm diameter apertures patterned on a PMMA layer using electron beam lithography (EBL) deposited on top of the SiC surface\cite{key-19}. We first measured the photoluminescence (PL) spectrum and second-order autocorrelation function $g^{2}(\tau)$ of the emision from the defects at room temperature and confirmed that they are single defects. We then studied their saturation behavior and photostability, showing that the fluorescence emission is very stable without any indication of photoblinking. By sampling the $V_{Si}$ defects on 100 implanted sites, we estimated that the efficiency of single silicon vacancy defects generation is about 41$\%$ and the conversion yield of implanted carbon ions into $V_{Si}$ defects is about 19$\%$. Finally, to confirm their origin further, we measured the PL spectrum and optically detected magnetic resonance (ODMR) signal of the $V_{Si}$ defects at low temperature (5 K).

\begin{figure*}
\includegraphics[scale=2]{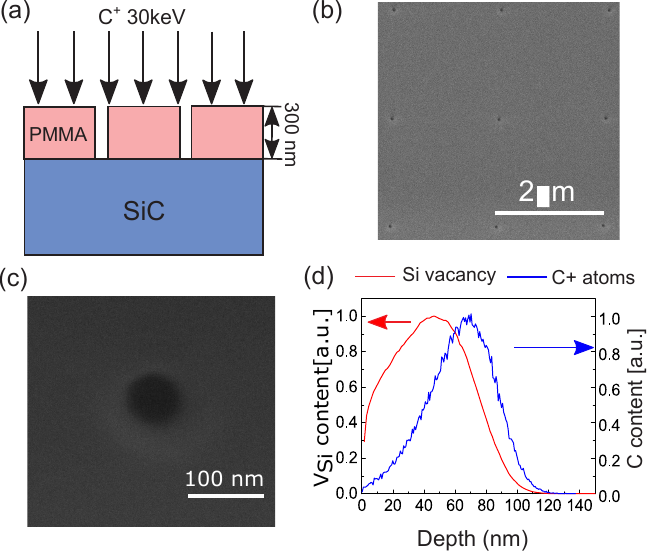}\caption{(a) The schematic of the carbon ion implantation through PMMA coating to generate single $V_{Si}$ defects array.(b) Scanning electron microscope (SEM) image of apertures array with 2$\mu m$ cell separation patterned on the surface of the PMMA. (c) Scanning electron microscope (SEM) image of an ~65 nm diameter aperture patterned on the surface of the PMMA. (d) The SRIM simulation of the depth profiles of the implanted carbon atoms (Blue) and the generated silicon vacancy (Red) in SiC at an energy of 30 keV.}

\label{Figure 1}
\end{figure*}

We first introduce the production steps for the defect centers. We started with a commercially available high-purity 4H-SiC epitaxy layer sample. This ensures a low background suitable for single-defect generation \cite{key-3,key-4}. As shown in Figure 1(a), after cleaning by acetone and isopropanol (IPA) in ultrasonic bath, a 300nm thick PMMA layer (A7:A4 = 5:3) was deposited on SiC surface by spin-coating \cite{key-19,key-20,key-21}. In the second step, an array of apertures with 2$\mu m$ cell separation and some long 10$\mu m$ wide strips used as position marks were generated using electron beam lithography (EBL) \cite{key-19}. After development step, a series of holes were generated on the PMMA layer. With scanning electron microscope (SEM), we can clearly see a hole of the size about 65 nm diameter (Figure 1(b) and 1(c)). Afterwards, 30keV $C^{+}$ ions were implanted with the fluence equal to 2.6 \texttimes{} 10\textsuperscript{11} $C^{+}/cm^{2}$ to generate $V_{Si}$ the defect array. The fluence corresponds to an average of around 8.6 carbon per aperture.  Using SRIM (Stopping and Range of Ions in Matter) software simulation, we found that more than $99\%$ of the 30keV carbon atoms can be blocked by the PMMA layer such that defect can only be implanted through the apertures. After the implantation, the PMMA layer on the sample was removed by ultrasonication in acetone. Finally, the sample was cleaned by ultrasonication in IPA. In order to avoid the generation of other types of PL defects, the sample was not annealled \cite{key-3}. Different from the generation of NV center in diamond, the absence of annealing steps largely simplified the production procedure.  In order to estimate the accuracy of the $V_{Si}$ defect position, we consider the ion straggling effect illustrated in Figure 1(d) which shows the depth profiles of the implanted carbon atoms and the generated Silicon vacancy at an energy of 30keV as simulated by SRIM. The average depth of the $V_{Si}$ defects is about 42nm and the longitudinal straggling is about 35nm.

After the production process, we characterized the emission properties of the defects. First, we studied the photoluminescence (PL) property of the emitters in a homemade confocal microscopy system \cite{key-4}. A 690nm continuous wave laser was used to excite $V_{Si}$ defects through a high N.A. oil objective lens (N.A.= 1.3, Nikon). The fluorescent photons from $V_{Si}$ defects were collected by the same objective lens and transmitted through a dichroic mirror (801nm). In order to suppress the background, the fluorescent photons were passed through a 75$\mu m$ diameter pinhole between two lenses followed by a 900nm long pass filter. The photons are then directly detected by two avalanche photodiodes (APDs) after a beam splitter in a Hanbury Brown and Twiss (HBT) setup. PL count map from a scanning area of $16\mu m \times 16\mu m$ is illustrated in Figure 2(a). As shown in this figure, the defect arrays can be seen clearly using the PL count map. The nearest two spots in Figure 2(a) have a cell separation distance of 2 $\mu$m which is consistent with the patterned PMMA apertures. The fluorescence spectrum of one of the spot (circled in Figure 2(a)) at room temperature is shown in Figure 2(b). Here the spectrum is cut by the 900nm long pass which is used to block the laser and oil fluorescence background. It agrees with the room temperature PL spectrum of $V_{Si}$ defect as measured in previous works \cite{key-3,key-4,key-22}.

\begin{figure}
\includegraphics{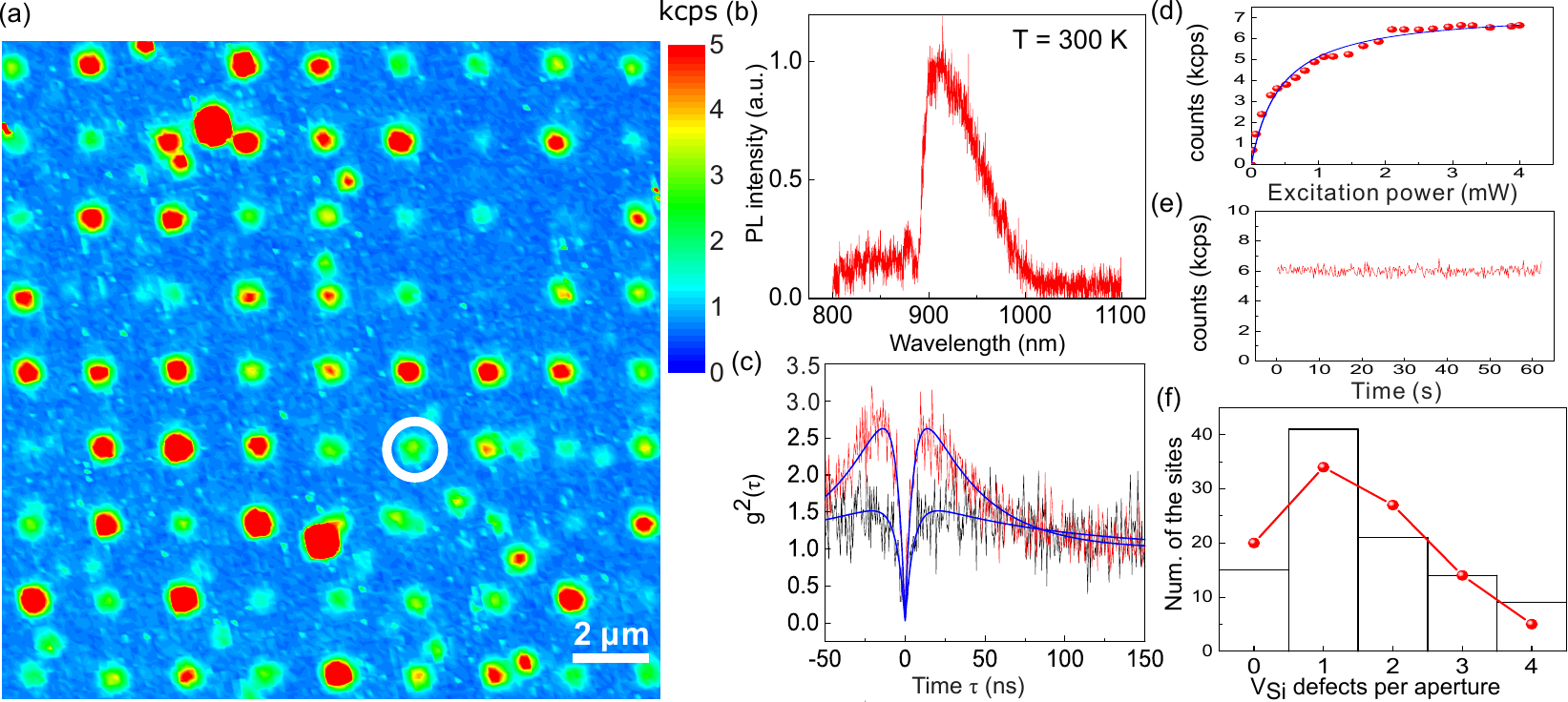}

\caption{(a) Confocal fluorescence image of an area of the sample with the implanted $V_{Si}$ defects array on the SiC surface. The circled one was the single $V_{Si}$ defect studied in this paper. The scale bar is 2$\mu m$. (b) The PL spectrum of the single $V_{Si}$ defect (circled in Figure 2(a)) at room temperature. (c) Second order correlation function measurement with excitation power of 0.5mW and 2mW, respectively. The blue lines show the fitting of the data with function introduced in the main text. (d) The PL count curve of the single $V_{Si}$ defect as a function of the excitation power. (e) The PL count of the single $V_{Si}$ defect as a function of time showing the photostability. Time bin is 100ms and excitation power is 2mW here. (f) The histogram of the number of $V_{Si}$ defects per aperture. The red line is the fit of the data by a Poisson distribution function. }

\label{Figure 2}
\end{figure}

In order to infer the number of the $V_{Si}$ defects associated with the luminescent spots, we measured the second order correlation function using the HBT setup. Since the saturation count for single $V_{Si}$ defect is about 7kcps, the signal to noise ratio (SNR) is low (about $4:1$), which will make the correlation function affected by the background noise. Here, the raw data $N(\tau)$ of the correlation functions is corrected using the function $g^{2}(\tau)=[N(\tau)-(1-\rho^{2})]/\rho^{2}$, where $\rho=s/(s+b)$, and s and b are the signal and background count \cite{key-3,key-6}, respectively. For the circled $V_{Si}$ defect in Figure 2(a), the value of $\rho$ is about $0.8$. Figure 2(c) shows the corrected correlation functions $g^{2}(\tau)$ of the circled $V_{Si}$ defect with 0.5mW (black line) and 2mW (red line) excitation power, respectively. The blue lines are the fit with the function $g^{2}(\tau)=1-(1+a)exp(-|\tau|/\tau_{1})+aexp(-|\tau|/\tau_{2})$, where $a$, $\tau_{1}$, $\tau_{2}$ are three fitting parameters. For the fitting function, we considered a three-level system with $\tau_{1}$ and $\tau_{2}$ are related to the emission and shelving state lifetimes \cite{key-23,key-24}. Fitting curves are shown in Figure 2(c) which agree with the single emitter model for both excitation powers. For 0.5mW excitation, we have $\tau_{1}=5.2\pm 0.5 ns$ and $\tau_{2}=89.1 \pm 8.5 ns$; for 2mW excitation, we have $\tau_{1}=5.3\pm 0.3 ns$ and $\tau_{2}=36.2 \pm 1.4 ns$.

We continue to show the saturation behavior of the single emitter. The PL saturation curve of the single $V_{Si}$ defect as a function of excitation power $P$ is presented in Figure 2(d). The blue line is the theoretical fitting function $I(P)=I_{s}/(1+P_{0}/P)$, where $I_{s}$ is the maximum count, and $P_{0}$ is the saturation power. The maximum count $I_{s}$ and the saturation power $P_{0}$ obtained from fitting are 7.4kcps and 0.43mW, respectively. Both values are in the same order with the former published results \cite{key-3,key-4}.  Under a similar setup, PL saturation count for single NV center in diamond can reach about $\sim 220$kcps. Given their similar lifetimes, it remains an open question whether the reduced count rate comes from the trapping in other dark states. Since it is not uncommon for a solid state emitter to have the blinking behavior, we next study its photostability which is much sought after in various quantum information applications. Figure 2(e) shows the PL count time trace with 100ms time bins under 2mW excitation power. This figure indicates that there is no photo-blinking in the time of 60s.

\begin{figure}
\includegraphics{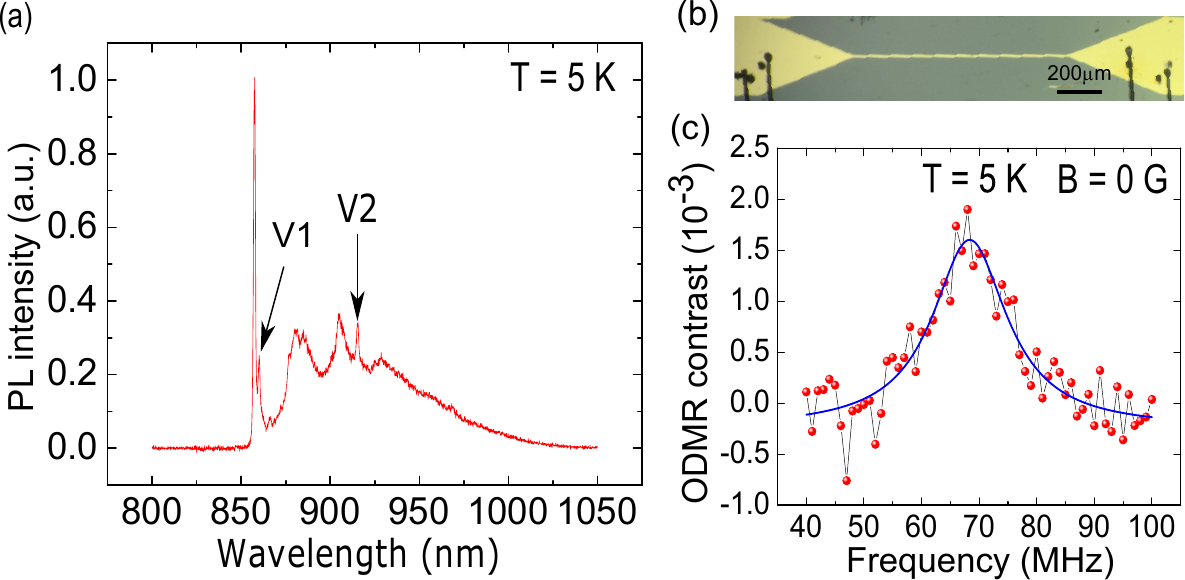}

\caption{(a) Low-temperature (5 K) PL spectrum of the $V_{Si}$ defects ensemble of the strip. The V1 and V2 ZPLs of $V_{Si}$ defects are observed at 860.1 nm and 915.3 nm, respectively. (b) Fabricated microwave stripline on sample surface. (c) Low-temperature (5 K) ODMR measurement of the $V_{Si}$ defects ensemble of the strip at zero magnetic. The solid curve is the Lorentzian fitting to the
spectrum. }

\label{Figure 3}
\end{figure}

Since the saturation count of the single $V_{Si}$ defect is only about 10kcps, high-efficiency generation of single $V_{Si}$ defect in a pre-defined position is critical for their integration with photonic cavity/waveguides to improve the emission and collection efficiency. In view of this, we give the statistics of the number of $V_{Si}$ defects per implanted aperture which is obtained through comparison between the measured the $g^{2}(0)$ values and PL intensity and the mean counts of a single $V_{Si}$ defect. Figure 2(f) shows the histogram of 100 apertures. The red line is the Poissonian distribution function fit of the data with a mean value equal to about 1.61. As can be seen, the number of single $V_{Si}$ defect is 41, corresponding to a single $V_{Si}$ defect generation efficiency of $41\%$. Since the implanted influence corresponds to about 8.6 carbon atom per aperture, the conversion yield of implanted carbon ions into $V_{Si}$ defects is about 19\%. The high conversion yield is comparable with that nitrogen implantation of the same energy used to generate the NV center in diamond \cite{key-19,key-25}.

After the generation of the defect, we focused on characterizing the origin of the implanted defect. The first experiment was to measure their PL spectrum in low temperature (5 K). $V_{Si}$ defect is a point defect consisting of a Silicon vacancy associated with adjacent four Carbon in 4H-SiC. There are two types of $V_{Si}$ defects in 4H-SiC: V1 and V2 centers. At room temperature, their PL spectrum wavelengths range from 800nm to 1100nm, which is in the near-infrared range. At crystal temperature,
there are two zero-phonon lines (ZPL), 861.4 and 916.3nm, corresponding
to V1 and V2 centers respectively \cite{key-4,key-16}. A spectrum of a defect ensemble is shown in Figure 3(a), from which we can see clearly the V1 and V2 peaks.

Lastly, we measured the continuous-wave optically detected magnetic resonance (ODMR) for the implanted $V_{Si}$ defects.  Only ODMR of V2 center is studied in this work. Its negatively charged ground state is a spin quartet state with $S=3/2$, exhibiting a zero-field splitting, D = 35MHz, which can be polarized by laser and controlled by microwave\cite{key-3}. The Hamiltonian of the V2 center can be expressed as
\begin{equation}
H=D(T)S_{z}^{2}+g\mu_{B}\vec{B} \cdot \vec{S}
\end{equation}
where g = 2.00 is the electron g factor, $\mu_{B}$ is the Bohr magneton, and $\vec{B}$ is the applied magnetic field.  ODMR measurement of $V_{Si}$ defect ensemble is shown in Figure 3(c). By doing Lorentzian fit of the ODMR data, we obtain a resonant frequency of 68.4 MHz, which is consistent with the zero field splitting, 2D ($\sim70 MHz$), of the V2 center in 4H-SiC \cite{key-3,key-4,key-16}. The slight difference might come from residual strains in different samples.

In summary, we experimentally demonstrate the generation of single silicon vacancy defect array in silicon carbide with a high efficiency. The successful generation of single silicon vacancy defect in a well-defined position around tens of nanometers may open up several immediate research possibilities. Firstly, a critical factor in quantum magnetometry with color center is the closeness of the sensing defect to the surface \cite{key-26}. This is because the dipolar magnetic fields decay as the third inverse power of the distance between the sensing spin and magnetic field target. We therefore need to have the defect centers close enough to the sample surface. The created shallow defects may find their application in nanoscale magnetometry, especially if it is combined with chemical etching \cite{key-27,key-28}. Secondly, it might serve as an efficient way to engineer spin-spin entanglement through dipole-dipole coupling \cite{key-29,key-30,key-31}. Again because of the coupling decays as distance between two defect spins with $d^{-3}$, it is much more favorable to generate two defects close to each other in the production procedure. The possibility to generate a few defects in the tens of nanometer scale as demonstrated here can be used towards this direction. Thirdly, to precisely couple the single emitter to a photonic crystal cavity \cite{key-12,key-32} , solid immersion lenses (SIL) \cite{key-33} or nano-pillar structures \cite{key-34}, one needs a precise location with accuracy in the order of tens of nanometers. The method demonstrated here may ease the coupling of emitter to fabricated cavity or waveguide. Finally, with modified parameters and possibly additional annealing steps, similar production of other types of single defects, such as divacancies \cite{key-1,key-2,key-6} and carbon antisite-vacancy pair defects \cite{key-8} may become possible in future.

\textbf{Materials and Method.}
\textbf{\textit{ODMR test.}} For ODMR measurement, a 4mW 690nm laser above the objective was used to excite the defects. The microwave signal (-10dbm) was generated by signal generator (Rohde $\&$ Schwarz, SMIQ 03E) and then gated by switch (ZASWA-2-50DR). After it is amplified by the amplifier (ZHL-20W-13+), it was sent into a Montana cyostation and then fed to a microwave strip line. As shown in Figure 3(b), the strip line with 20$\mu m$ width was fabricated on the sample surface by standard lithography method. The whole experiment was synchronized by pulse blaster (Spincore, PBESR-PRO-500). To decrease the fluctuation noise, ODMR scans from 40Mhz to 100Mhz were conducted six times and then the scan results are averaged. For one point in each scan, the microwave signal was gated on and off with 2.8ms duration and repeated 20000 times \cite{key-16}. The photon counts in each on and off microwave period was recorded by DAQ card (NI 6343) triggered by pulse blaster. Final ODMR contrast was calculated by  $\Delta PL=({\sum N(off)-\sum N(on)})/{\sum N(on)}$. According to the Lorentz fit of the data, we obtained the resonant frequency of 68.4 MHz, which was consistent with the zero field splitting, 2D (70 MHz), of the V2 center in 4H-SiC. \cite{key-3,key-4,key-17}

\begin{acknowledgement}
We thank the discussion with Phani Kumar and Abdullah Rasmita. We acknowledges the support from the Singapore National Research Foundation through a Singapore 2015 NRF fellowship grant (NRF-NRFF2015-03) and its Competitive Research Program (CRP Award No. NRF-CRP14-2014-02), and a start-up grant (M4081441) from Nanyang Technological University.

\end{acknowledgement}

\end{document}